\documentclass[referee]{aa} 

\usepackage{graphicx}
\usepackage{txfonts}
\usepackage{rotating}

\def\HI{\ion{H}{I}}          

\begin{document}

   \title{Orbital phase resolved spectroscopy of 4U1538$-$52 with MAXI}

   \subtitle{}

   \author{J. J. Rodes-Roca
          \inst{1,2,3}\fnmsep\thanks{JJRR thanks the Matsumae
          International Foundation Research Fellowship}
          \and
          T. Mihara\inst{3}
          \and
          S. Nakahira\inst{4}
          \and
          J. M. Torrej\'on\inst{1,2}
          \and
          \'A. Gim\'enez-Garc\'{\i}a\inst{1,2,5}
          \and
          G. Bernab\'eu\inst{1,2}
          }

   \institute{Dept. of Physics, Systems Engineering and Sign Theory,
University of Alicante, 03080 Alicante, Spain\\
              \email{rodes@dfists.ua.es}
         \and
             University Institute of Physics Applied to Sciences and
Technologies, University of Alicante, 03080 Alicante, Spain\\
         \and
             MAXI team, Institute of Physical and Chemical Research
(RIKEN), 2-1 Hirosawa, Wako, Saitama 351-0198, Japan\\
             \email{tmihara@riken.jp}
         \and
             ISS Science Project Office, Institute of Space and Astronautical
Science (ISAS), Japan Aerospace Exploration Agency (JAXA),
2-1-1 Sengen, Tsukuba, Ibaraki 305-8505, Japan\\
         \and
             School of Physics, Faculty of Science, Monash University,
Clayton, Victoria 3800, Australia\\
             }

   \date{XX-XX-XX; XX-XX-XX}

 
  \abstract
   {4U 1538$-$52, an absorbed high mass X-ray binary with an orbital
    period of $\sim$3.73 days, shows moderate orbital intensity
    modulations with a low level of counts during the eclipse. Several
    models have been proposed to explain the accretion at different
    orbital phases by a spherically symmetric stellar wind from
    the companion.}
   {The aim of this work is to study both the light curve
   and orbital phase spectroscopy of this source in the long term. Particularly,
   the folded light curve and the changes of the spectral parameters
   with orbital phase to analyse the stellar wind of QV Nor, the
   mass donor of this binary system.}
   {We used all the observations made from the Gas Slit Camera
   on board \emph{MAXI} of 4U 1538$-$52 covering many orbits
   continuously. We obtained the good interval times for every
   orbital phase range which were the input to extract our data.
   We estimated the orbital period of the system and then folded
   the light curves and we fitted the X-ray spectra with
   the same model for every orbital phase spectrum.
   We also extracted the averaged spectrum of all the
   \emph{MAXI} data available.}
   {The \emph{MAXI} spectra in the 2--20 keV energy range were
   fitted with an absorbed Comptonization of cool photons on
   hot electrons.
   We found a strong orbital dependence of the absorption
   column density but neither the fluorescence iron emission line
   nor low energy excess were needed to fit the \emph{MAXI} spectra.
   The variation of the spectral parameters over the binary
   orbit were used to examine the mode of accretion onto the
   neutron star in 4U 1538$-$52. We deduce a best value of
   $\dot{M}/v_\infty=0.65\times 10^{-9}$ $M_{\odot} \, yr^{-1}/(km \, s^{-1})$
   for QV Nor.}
   {}


   \maketitle
%

\section{Introduction}     \label{sec:introduction}

\object{4U 1538$-$52} is a high mass X-ray binary (HMXB) pulsar with a B-type
supergiant companion, QV Nor. The orbital period of the binary system is
$\sim$3.73 days (\cite{1977MNRAS.181P..73D}; 
\cite{2000ApJ...542L.131C}; 
\cite{2006JApA...27..411M}), 
and the magnetised neutron star has a spin period of
$\sim$529 s (\cite{1977MNRAS.181P..73D}; 
\cite{1977ApJ...216L..11B}). 
The distance to the source
was estimated to be between 4.0 and 7.4 kpc
(\cite{1978ApJ...225L..63C}; 
\cite{1979A&A....71L..17I}; 
\cite{1992MNRAS.256..631R}), 
and we use its latest estimation it in this work, i.e.,
6.4 kpc.
Although an orbital low eccentricity was estimated to be
$e\sim$0.08, (\cite{1993A&A...276...52C}), 
a higher value of $e\sim 0.17$ suggested an elliptical orbit
for this system (\cite{2000ApJ...542L.131C}; 
\cite{2006JApA...27..411M}). 
Nevertheless, \citet{2011ApJ...730...25R} 
argued that a circular orbit and an eccentric orbit cannot
be distinguished.
The X-ray luminosity is minimum during the eclipse which lasts
$\sim$0.6 days (\cite{1977ApJ...216L..11B}). 

Both timing and spectral properties of 4U 1538$-$52 have been
studied with some X-ray observatories in different energy
bands, such as \emph{Tenma} (\cite{1987ApJ...314..619M}),
\emph{EXOSAT} (\cite{1992ApJ...401..685R}), 
\emph{Ginga} \citep{1995ApJ...444..405B}, 
\emph{RXTE} (\cite{2000ApJ...542L.131C}), 
\emph{BeppoSAX} (\cite{2001ApJ...562..950R}), 
\emph{XMM-Newton} (\cite{2011A&A...526A..64R}), 
\emph{INTEGRAL} \citep{2013ApJ...777...61H}, and 
\emph{Suzaku} \citep{2014ApJ...792...14H}. 
The X-ray characteristics derived from these analyses suggest
that the mass accretion onto the neutron star is consistent
with a spherically symmetric stellar wind from the companion
star (\cite{2006JApA...27..411M}). 
The X-ray spectrum based on a broad band energy range has
been described by different absorbed power-law relations,
modified by a high-energy cutoff including a fluorescence
iron line at 6.4 keV, whenever present, and with cyclotron
resonant scattering features at $\sim$21 keV
(\cite{1990ApJ...353..274C}; \cite{2001ApJ...562..950R})
and at $\sim$47 keV (\cite{2009A&A...508..395R}). 
A soft excess component is found in the X-ray spectrum from
\emph{BeppoSAX} and \emph{XMM-Newton} observations
(\cite{2001ApJ...562..950R}; \cite{2011A&A...526A..64R}).
Similar to other wind-fed pulsars, the X-ray flux
changes while the neutron star is moving in its orbit
around the circumstellar environment of the companion star.
Therefore, the variability is associated with
the mass-loss rate by stellar wind captured by
the gravitational field of the neutron star ($\dot{M}$).
It also depends on the distance between both components
and the stellar wind velocity, mainly. The estimated
X-ray luminosity is $\sim (2.6-9.1)\times 10^{36}$ erg\,s$^{-1}$
in the 3--100 keV range, assuming an isotropic emission
and a distance of 6.4 kpc (\cite{1992MNRAS.256..631R}).
Thus, the analysis of the X-ray spectrum from the
neutron star at different orbital phases provides us
the variability of the model parameters we can use
to compare with accretion models.

The Monitor of All Sky X-ray Image (\emph{MAXI})
presents both all sky coverage and moderate energy
resolution which gives us the possibility to investigate
the orbital light-curves and the orbital phase resolved
spectra of 4U 1538$-$52. In this paper, we used \emph{MAXI}
data to study the light curves, the orbital phase averaged
spectrum, and the orbital phase resolved spectra of 4U 1538$-$52
in the 2--20 keV energy range.
Contrary to previous studies which focus on specific
orbital phases at specific times, the more than five years of
\emph{MAXI} data, analysed in this work tend to smear out
short time scale variations and give more weight to long
term accretion structures.
We describe the
observations and the data reduction in Sect.~\ref{sec:observations},
present the timing analysis in Sect.~\ref{sec:timing},
the spectral analysis in Sect.~\ref{sec:spectra}
and in Sect.~\ref{sec:conclusions} we summarise our results.
All uncertainties are hereafter given at the 90\%
($\Delta\chi^2\!=\!2.71$) confidence limit, unless otherwise
specified.

\section{Observations}    \label{sec:observations}

\emph{MAXI} is the first astronomical mission on the International
Space Station (ISS; \cite{2009PASJ...61..999M}). 
\emph{MAXI} attached to the Japanese Experiment Module--Exposed Facility (JEM--EF).
It consists in two types of X-ray slit cameras. The
main X-ray camera is the Gas Slit Camera (GSC; \cite{2011PASJ...63S.623M})
operating in the 2--20 keV energy range. The second one is the
Solid-state Slit Camera (SSC; \cite{2011PASJ...63..397T})
operating in the 0.7--7 keV energy range. The in-orbit
performance of GSC and SSC is presented in
\citet{2011PASJ...63S.635S} 
and \citet{2010PASJ...62.1371T}, respectively. 

Every hour and a half \emph{MAXI} covers almost the entire sky
per ISS orbit. Therefore, the source is observed around 1 ks per
day by \emph{MAXI}. We accumulated the exposure time to extract
the spectra with enough level of counts.

\section{Timing analysis}    \label{sec:timing}

\begin{figure}[hbtp]
  \centering
  \includegraphics[angle=-90,width=\columnwidth]{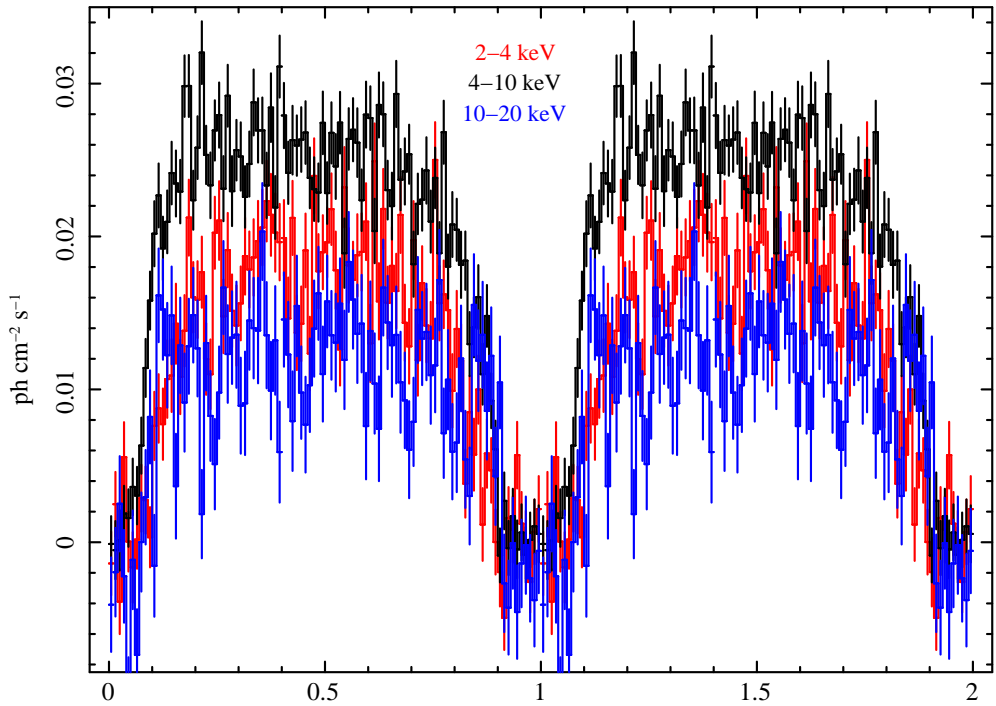}  
  \caption[]{\label{fig:lcurve} %
    \emph{MAXI} orbital phase resolved background subtracted light curves
    of 4U 1538$-$52
    in 2--4 keV, 4--10 keV, and
    10--20 keV energy ranges using orbital parameters
    reported by \citet{2006JApA...27..411M}, and folded it
    relative to the mid-eclipse phase.
  }
\end{figure}

We have first obtained the \emph{MAXI/GSC} 1-orbit light-curve\footnote{\emph{MAXI}
homepage \href{http://maxi.riken.jp}{\url{http://maxi.riken.jp}}} of
4U 1538$-$52 from MJD 55\,058 to MJD 56\,821 (i.e. nearly 5 years) in the 2--20 keV
energy band to estimate the orbital period of the binary system.
Then, we have searched for a period assuming a sinusoidal signal
and the error was obtained using the relations derived by
\citet{1987A&A...180..275L}. 
Our result was $P_{\mathrm{orb}}\!=\! 3.7285\pm 0.0006$ days
which is consistent with the value given by \citet{2006JApA...27..411M}
($P_{\mathrm{orb}}\!=\! 3.728382\pm 0.000011$ days). 
Then we have folded the light curves with the best orbital period
to obtain energy resolved orbital intensity profiles. The orbital
phase reference is taken from \citet{2006JApA...27..411M},
with the phase zero corresponding to mid-eclipse
in Fig.~\ref{fig:lcurve}.


The \emph{MAXI/GSC} data used in our analysis were extracted
by the \emph{MAXI} on-demand process using a 1.6 degrees radius
circular region centred at the X-ray source position with an
annulus background region where we excluded other bright sources
in the field.
In Fig.~\ref{fig:lcurve} we show the background subtracted light curves for the
\emph{MAXI/GSC} camera, folded at the orbital period of $\sim$3.73 days,
in the energy ranges 2--4 keV, 4--10 keV, and 10--20 keV.
We note that the X-ray flux during the eclipse is compatible
with no counts in all the energy bands taking the uncertainties into
account.

\section{Spectral analysis}    \label{sec:spectra}

\subsection{Orbital phase averaged spectrum}

We have extracted the orbital phase averaged spectrum of 4U 1538$-$52
with \emph{MAXI/GSC} for the same observation duration using the
\emph{MAXI} on-demand processing\footnote{\href{http://maxi.riken.jp/mxondem}{\url{http://maxi.riken.jp/mxondem}}}.
For spectral analysis we used the \textsc{XSPEC} version 12.8.1
(\cite{1996ASPC..101...17A}) fitting package, 
released as a part of \textsc{XANADU} in the HEASoft tools.
We used \textsc{FTOOL grppha} to group the raw spectra
as 2-133 channels by 4, 134-181 channels by 8, and
182-1199 channels by 12  so that all the spectral bins were Gaussian
distributed. We tested both phenomenological and physical models
commonly applied to accreting X-ray pulsars.

\begin{figure}[hbtp]
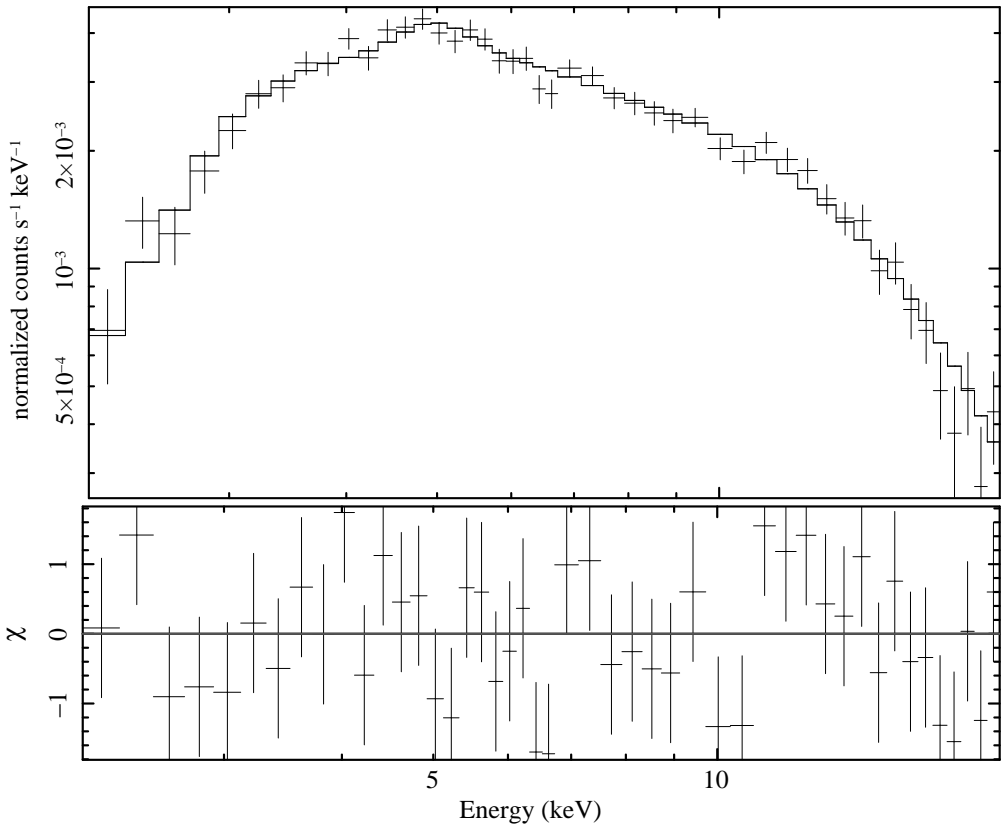

  \centering
  \includegraphics[angle=-90,width=\columnwidth]{tbabscompstcyclabs-bin}  
  \includegraphics[angle=-90,width=\columnwidth]{tbabsnthcompcyclabs-bin}  
  \caption[]{\label{fig:average} %
    Orbital phase averaged spectrum of 4U 1538$-$52
    in 2--20 keV band and X-ray model continuum modified by the
    fundamental CRSF. 
    \emph{Top panel}: Data and absorbed Comptonization of
    cool photons on hot electrons.
    \emph{Bottom panel} shows the residuals between the spectrum
    and the model.
  }
\end{figure}

Comptonization models attempt to provide
a better physical description of the underlying emission
mechanisms. We fitted the 2--20 keV energy spectrum
with several successful models.
We have chosen \textsc{CompST} (in \textsc{XSPEC}
terminology; \cite{1980A&A....86..121S}) 
as the representative model, i.e., a Comptonization of cool photons
on hot electrons modified by an absorbing column along our line of sight.
As \emph{MAXI/GSC} does not have good enough sensitivity to
detect the fluorescence iron
emission line at $\sim$6.4 keV we have not needed to include it
in our fits. The absorption column $N_H$ fitted with other models were
consistent, within uncertainties, and we will use the \textsc{CompST}
values in the rest of this work.
The soft excess at low energies (0.1--1 keV) presents
in 4U 1538$-$52 and other HMXBs has not been considered here because
it lies outside the energy range of the \emph{MAXI/GSC} spectrum.
Nevertheless, the well known fundamental cyclotron resonant
scattering feature (CRSF) at $\sim$21 keV modifies the X-ray continuum
above 16 keV (see references in Sect.~\ref{sec:introduction}).
Therefore, we included in our models the CRSF
fixing its parameters with the values obtained by
\citet{2009A&A...508..395R}. 
We used the absorption cross sections from
\citet{1996ApJ...465..487V}, 
and the abundances are set to those of \citet{2000ApJ...542..914W}.
The fitted parameters for the continuum model is listed
in Table~\ref{tab:average}.
In Fig.~\ref{fig:average} we plot our data together with the
absorbed \textsc{CompST} best-fit
model, and residuals of the fit as the difference between the
observed flux and model flux divided by the uncertainty of
the observed flux.

%
\begin{table}
\caption{Fitted parameters for the \emph{MAXI} spectra in
Fig.~\ref{fig:average}.}             
\label{tab:average}      
\centering                          
\begin{tabular}{l c}        
\hline\hline                 
Parameter & \textsc{CompST} \\    
\hline                        
   $N_H$ ($\times 10^{22}$ cm$^{-2}$) & 4.9$\pm$1.4 \\      
   $kT$ (keV) & 4.7$^{+0.7}_{-0.5}$ \\
   $\tau$ & 17.8$^{+2.4}_{-2.2}$ \\
   Norm ($\times 10^{-2}$) & 2.7$\pm$0.6 \\
   Flux (2--20 keV) & 6.5$^{+0.3}_{-0.2}$ \\
   $\chi^2$/dof & 40.6/43 \\
\hline                                   
\end{tabular}
\tablefoot{A CRSF was included in the model (see text). Unabsorbed
flux in units of ($\times 10^{-10}$) erg\,cm$^{-2}$\,s$^{-1}$.}
\end{table}
%

As the total Galactic
\HI\ column density in the line of sight of 4U 1538$-$52 is
in $(9.14-9.70)\times 10^{21}$ atoms\,cm$^{-2}$ range
(\cite{1990ARA&A..28..215D}, 
\cite{2005A&A...440..775K}), 
the neutron star is moving almost all the orbit in
additional absorbing material related to the stellar wind
of the companion. Therefore, we studied its variability
around the orbit extracting
phase resolved spectra in the next Section.

\subsection{Orbital phase resolved spectra}

Previous works in this sense have been carried out
for only one or two orbits (\cite{1987ApJ...314..619M}, 
\cite{2001ApJ...562..950R}, 
\cite{2006JApA...27..411M}, 
\cite{hdl.handle.net/10045/13227}) 
or for a shorter orbital phase (\cite{2011A&A...526A..64R}). 
We have obtained orbital phase resolved spectra of the
HMXB pulsar 4U 1538$-$52 accumulating the 60 s
duration scans into six $\approx$0.5 day orbital
phase bins outside of eclipse
(\cite{2012PASJ...64...13N}, 
\cite{2013A&A...554A..37D}, 
\cite{2014MNRAS.441.2539I}). 

\begin{figure}[hbtp]
  \centering
  \includegraphics[angle=-90,width=\columnwidth]{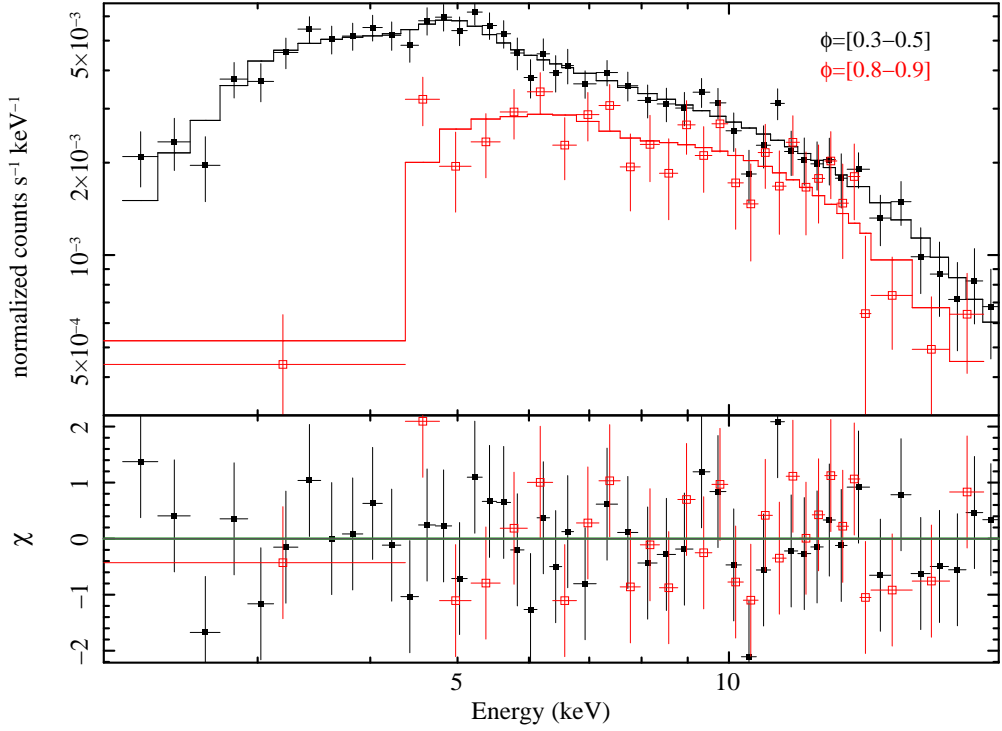}  
  \caption[]{\label{fig:resolved} %
    Orbital phase resolved spectra of 4U 1538$-$52
    in 2--20 keV band.
    \emph{Top panel}: Selected spectra and absorbed \textsc{CompST}
     best-fit model.
    \emph{Bottom panel} shows the residuals for the model.
  }
\end{figure}

In this analysis, we have fitted the orbital phase resolved
spectra with the same model we used in the orbital phase
averaged spectrum (see previous Section). Although we also
extracted the \emph{MAXI/SSC} spectra, the time coverage of
the SSC was too low to study spectral and flux changes. We
therefore concentrate on the GSC data. For all the spectral models,
we kept the parameters free,
the orbital phase resolved spectra were fitted from energy
range 2 keV to 20 keV.
The range of the $\chi^2_\nu$ values
was 0.7--1.1 for all fits. The data together with the
absorbed \textsc{CompST} best-fit model,
and residuals as the difference between observed flux and model
flux divided by the uncertainty of the observed flux,
are shown in Fig.~\ref{fig:resolved}. We obtained six
orbital phases to keep a good S/N ratio. In Fig.~\ref{fig:resolved}
we plot the highest and
lowest flux spectra and two sets of residuals.

\begin{figure}[hbtp]
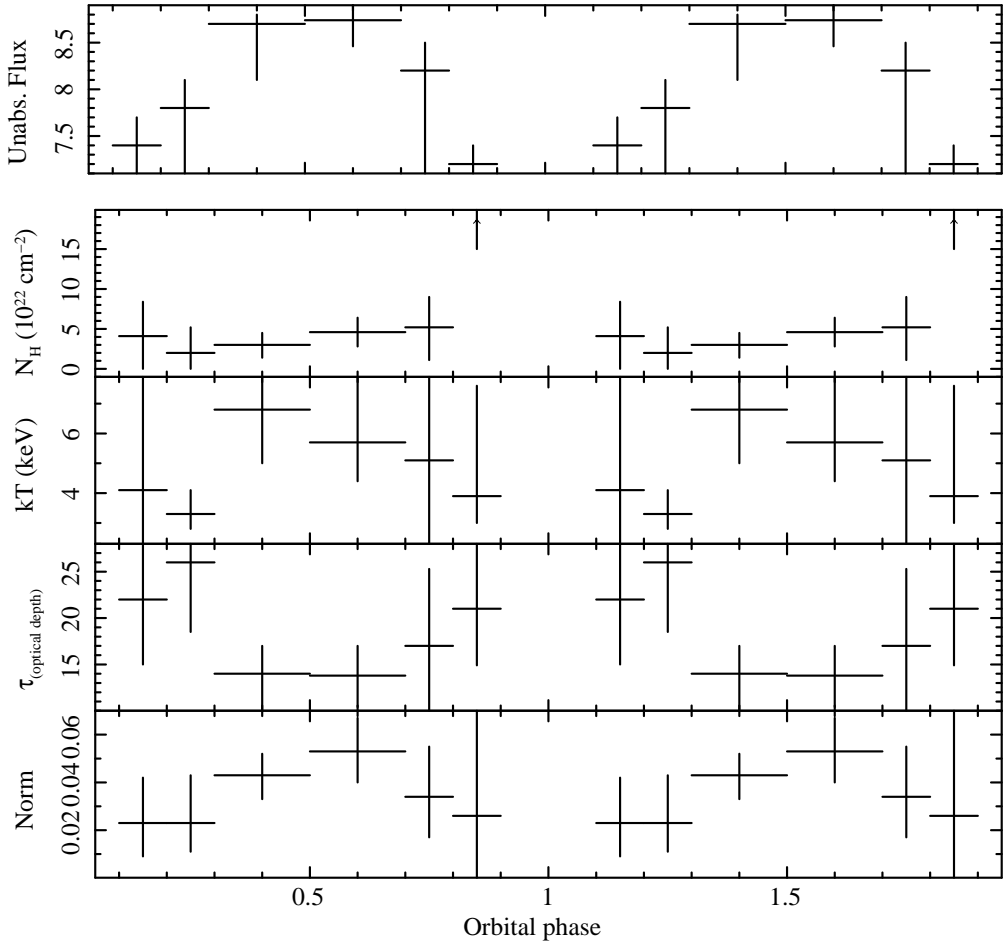

  \centering
  \includegraphics[angle=-90,width=\columnwidth]{Flux-compst-bin2}  
  \includegraphics[angle=-90,width=\columnwidth]{parameters-compst2-bin2}  
  \caption[]{\label{fig:orbitcompst} %
    Orbital phase changes in the free spectral parameters fitted
    in 2--20 keV band. Spectral parameters associated with the absorbed Comptonization of cool
    photons by hot electrons (\emph{compST}) model. Unabsorbed flux in units
    of $10^{-10}$ erg\,cm$^{-2}$\,s$^{-1}$.
  }
\end{figure}

The variability of the spectral parameters for the
absorbed thermal Comptonization with a spherical geometry
(\textsc{CompST} in \textsc{XSPEC}) is shown in
Figure~\ref{fig:orbitcompst}.
The Comptonization parameter, $y= k\,T\,\tau^2/(m_{\rm e}\,c^2)$,
determines the efficiency of the Comptonization process
(\cite{1994ApJ...434..570T}, 
\cite{2008MNRAS.389..301P}) 
and we derived its value from the six orbital phase ranges
(see Table~\ref{tab:accretion}). This indicates an efficient process
which corresponds to a moderate accretion rate, being
higher in the orbital phase ranges just before and after
the eclipse than in the mid orbital phase ranges. Moreover,
it is consistent with similar observations of Vela X--1
performed by \emph{MAXI} (\cite{2013A&A...554A..37D}).
The unabsorbed flux along the orbit has no significant evolution,
$(7.2-8.8)\times 10^{10}$ erg\,cm$^{-2}$\,s$^{-1}$ in the
fitting energy range (2-20) keV,
showing that the accretion rate is quite stable, in the phase
bins considered.

%
\begin{table}
\caption{X-ray luminosity and Comptonization parameter}             
\label{tab:accretion}      
\centering                          
\begin{tabular}{c c c}        
\hline\hline                 
Orbital Phase & $L_X$ ($10^{36}$ erg\,s$^{-1}$) & $y$ \\    
\hline                        
   $[0.1-0.2]$ & 3.6 & 3.8 \\      
   $[0.2-0.3]$ & 3.9 & 4.9 \\      
   $[0.3-0.5]$ & 4.1 & 2.6 \\
   $[0.5-0.7]$ & 4.0 & 2.2 \\
   $[0.7-0.8]$ & 3.8 & 3.0 \\
   $[0.8-0.9]$ & 3.3 & 3.7 \\ 
\hline                                   
\end{tabular}
\end{table}
%

\subsection{The stellar wind in 4U 1538$-$52}

\begin{figure}[hbtp]
  \centering
  \includegraphics[angle=0,width=\columnwidth]{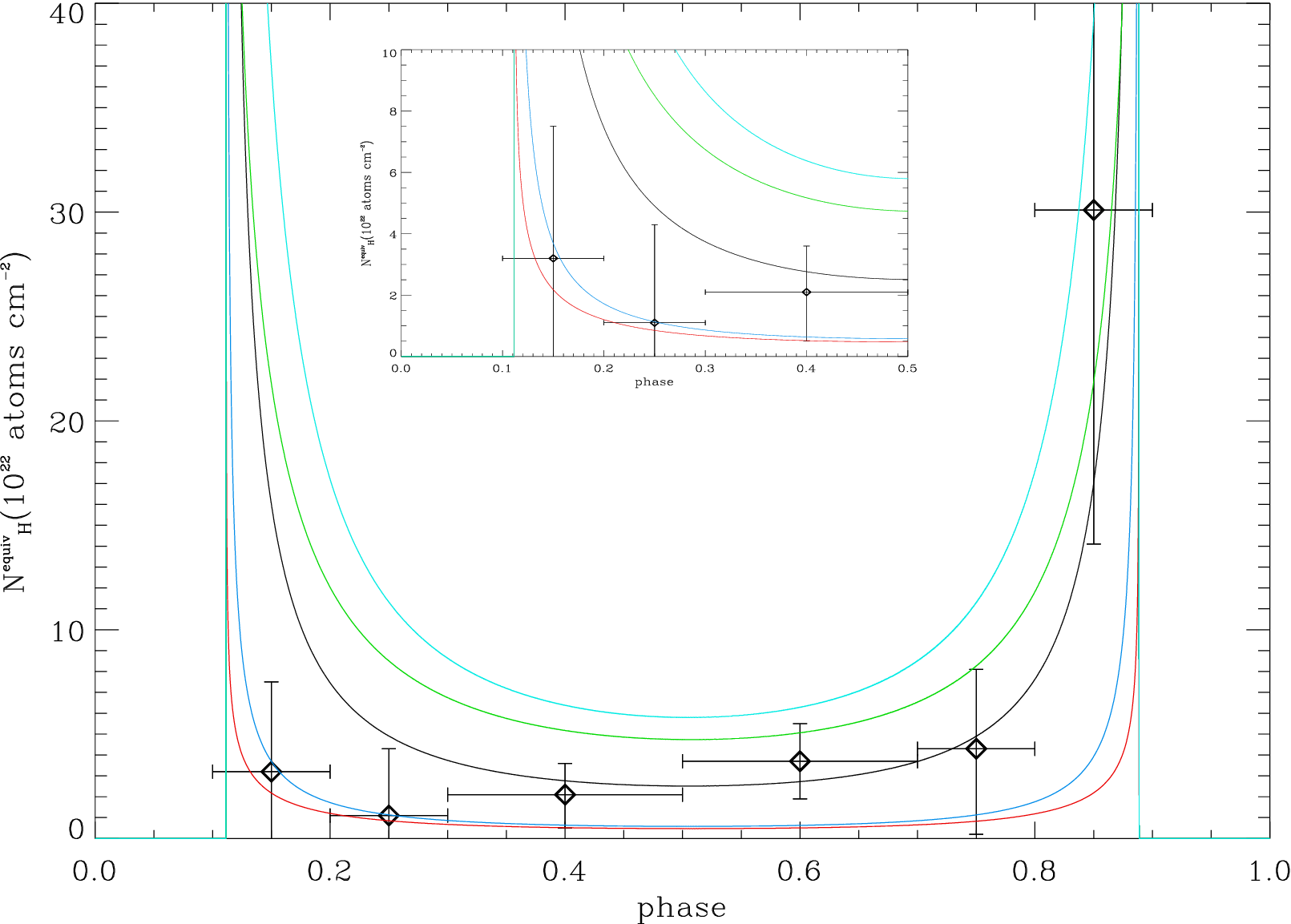}  
  \caption[]{\label{fig:wind} %
    Variation of column density versus orbital phase. Red, blue, black, green,
    and turquoise lines represent the wind model for different values of $\beta$
    and $\dot{M}/v_\infty$. See text for details. 
  }
\end{figure}

One way to study the stellar wind in HMXBs is in terms of the
variability of the equivalent hydrogen column density throughout
the binary orbit. In fact, a simple spherically
symmetric stellar wind model may describe the observed orbital
dependence of the column density for certain range of the
orbital inclination (\cite{1994ApJ...422..336C}, 
\cite{2006JApA...27..411M}, 
\cite{2008xru..confE..56R}). 
In order to obtain an estimation of the stellar mass-loss
rate, the equivalent absorption column was derived by
integrating the wind density along the line of sight to the
X-ray source, i.e., combining the simple spherically symmetric
wind model and conservation of mass equations. Then,
the hydrogen number density at distance $r$ from the donor,
$n_\text{H}(r)$, can also be estimated through the following
relationship:

\begin{equation}
n_\text{H}(r) = \frac{X_\text{H}\,\dot{M}}{m_\text{H}\,v_\infty\,(1- 
R_{\star}/r)^{\beta}\,4\,\pi \,r^2}
\label{eq:nhphase}
\end{equation}
with $m_\text{H}$ the hydrogen atom mass, $X_H$ the hydrogen mass 
fraction, $\dot{M}$ the mass loss rate,
$v_\infty$ the terminal velocity of the wind in the range 1\,400--2\,800
km\,s$^{-1}$ \citep{1982ApJ...263..723A}, 
where $r = 23.4 \, R_\odot$ represents the binary separation
\citep{2000ApJ...542L.131C}, 
the parameter $\beta = 0.8-1.2$ is the velocity gradient
for an OB supergiant \citep{1986ApJ...311..701F}, 
and $R_{\star} = 17.2 \, R_\odot$ \citep{1992MNRAS.256..631R} 
the radius of the donor.
The resultant $N_\text{H}$ 
arises from the integration of $n_H(r)$ along the line of sight.
In Fig.~\ref{fig:wind} we show the equivalent hydrogen column
density measured with \emph{MAXI} as a function of the
orbital phase.

The lines represent the model calculations
of the absorption column density for different values of $\beta$ and
$\dot{M}\, \, (M_\odot \, {\rm yr^{-1}})/v_\infty \, \, (\text{km\, s}^{-1})$,
in units of $10^{-9}$. The red
line corresponds to $\beta = 0.8$ and
$\dot{M}/v_\infty = 0.3$;
the blue line corresponds to $\beta = 1.2$ and
$\dot{M}/v_\infty = 0.3$;
the green line corresponds to $\beta = 0.8$ and
$\dot{M}/v_\infty = 3.0$;
and the turquoise line corresponds to $\beta = 1.2$ and
$\dot{M}/v_\infty = 3.0$.
Our best least-squares fit is the black line which
corresponds to $\beta = 1.2$ and $\dot{M}/v_\infty = 1.3$.
The value of the other orbital parameters have been taken from
\citet{2000ApJ...542L.131C} 
and \citet{2006JApA...27..411M}. 
As can be seen, our data can not discriminate between different values of $\beta$.
We also note that our estimations of $\dot{M}/v_\infty$
did not change very much with the eccentricity and orbital inclination.
None of the fits could describe all the experimental data. Consequently, we looked for
the upper and lower values of a better fit.
From the eclipse egress to the mid-orbit,
the models with a lower $\dot{M}/v_\infty$ fit the data well (red and blue lines), but from
the mid-orbit to the eclipse ingress, a higher value of $\dot{M}/v_\infty$ (green and turquoise lines)
is required. This fact suggest that the neutron star is crossing an
overdense region of the stellar wind and/or some trailing
accreting material passing through the line of sight.
Moreover, hydrodynamical simulations show that the interaction between the
gravitational field of the neutron star, the radiation field of the X-ray
source and the stellar wind, may lead to the formation of large structures
of higher density compared to the unperturbed stellar wind
\citep{1990ApJ...356..591B}. 
These structures are formed mainly in the trail of the
neutron star. Consequently, they act as X-ray absorbers predominantly in the
second half of the orbit. In Vela X$-$1, an excess of absorption at late orbital
phases is usually observed. This excess has been successfully modelled by means
of the above-mentioned hydrodynamical simulations
\citep{2012A&A...547A..20M}. 
The fact that Vela X$-$1 and 4U 1538$-$522 are similar
systems, since both have a donor of spectral type B0-0.5 I and very short
orbital periods, imply that these structures are very likely to be present
in 4U 1538$-$522.

Assuming that the eclipse egress ($\phi=[0.1-0.3]$) represents the unperturbed wind,
the best value of $\dot{M}/v_\infty$ would be in the range
$(0.3-1.0)\times 10^{-9}$ $M_{\odot} \, yr^{-1}/(km \, s^{-1})$. The median
value of this range, namely $0.65\times 10^{-9}$ $M_{\odot} \, yr^{-1}/(km \, s^{-1})$,
is in excellent agreement with that obtained by \citet{1994ApJ...422..336C}.
For the terminal velocities reported by \citet{1982ApJ...263..723A}
this would translate into a mass loss of $(0.4-2.8)\times 10^{-6} M_{\odot} \, yr^{-1}$
for QV Nor.

\citet{1990A&A...231..134N} investigated the dependence
of $\dot{M}$ on the stellar fundamental parameters mass $M$,
radius $R$, and luminosity $L$ for a large sample of stars.
They presented a simple parametrization that gave a good
description of observed mass-loss rates over the whole
Hertzsprung-Russell diagram. We noted that their formula has
different numbers in the abstract than in the text which
we think it is the correct one. Therefore, we used the following
expression:

\begin{equation}
\dot{M} = 9.55\times 10^{-15} \left( \frac{L}{L_\odot} \right)^{1.24}
\left( \frac{M}{M_\odot} \right)^{0.16}
\left( \frac{R}{R_\odot} \right)^{0.81} \, \, M_\odot \, {\rm yr^{-1}}
\, .
\label{eq:mdot}
\end{equation}

The astrophysical data for QV Nor are
$\log(L/L_\odot) = 5.21\pm 0.13$, $R/R_\odot = 17.2\pm 1.0$,
and $M/M_\odot = 20\pm 4$ \citep{1992MNRAS.256..631R}.
By substituting these values in the Eq.~\ref{eq:mdot},
the mass-loss rate was estimated as
$\dot{M} = 4.5\times 10^{-7} \, \, M_\odot \, {\rm yr^{-1}}$,
in agreement with the smaller value obtained by considering
a spherically symmetric wind.

Our values agree also with the Abbott's correlation
between mass-loss rate and luminosity, $\dot{M} = (0.03-1.0)\times 10^{-6}$
$M_\odot$\,yr$^{-1}$ \citep{1982ApJ...263..723A}. 
Finally, \citet{1994ApJ...422..336C}, using a {\it Ginga} pointed observation, obtained a range $\dot{M} = (0.9-1.9)\times 10^{-6}$
$M_\odot$\,yr$^{-1}$, for the eclipse egress data, compatible with our estimation. 
Therefore, combining all the previous estimates we can deduce a mass loss range for
4U 1538$-$52 of $\dot{M} = (1.2\pm 0.7)\times 10^{-6}$
$M_\odot$\,yr$^{-1}$.
These values would point to terminal wind velocities larger than 2\,000
km\,s$^{-1}$,
closer to the upper limit (2\,800 km\,s$^{-1}$) of the
range derived by \citet{1982ApJ...263..723A}. 

\section{Conclusions} \label{sec:conclusions}

From the analysis of the \emph{MAXI/GSC} light curve, we have estimated
the orbital period of the binary system,
$P_{\mathrm{orb}}\!=\! 3.7285\pm 0.0006$ days, being in agreement with
the best value derived by \citet{2006JApA...27..411M}.
From the unabsorbed flux, the X-ray luminosity in the 2--20 keV
energy band was found, $L_X = (3.8 \pm 0.5)\times 10^{36}$ erg\,s$^{-1}$.
We have investigated the long term orbital variation of spectral
parameters doing an orbital phase resolved spectroscopy of
4U 1538$-$52 with \emph{MAXI}, nearly five years of data.
A higher value of $\dot{M}/v_\infty$ is
clearly needed to fit the absorption column from orbital
phases 0.4 into eclipse ingress.
Using the $N_H$ values up to orbital phase 0.3, which we
consider the unperturbed wind, we measured the orbital dependence
of the column density in the system and of the intrinsic source
X-ray flux, and found that it was consistent on average with a spherically
symmetric stellar wind from the optical counterpart of the binary
system. We have estimated the mass-loss rate range of the
early-type B supergiant star QV Nor to be $\dot{M} = (1.2\pm 0.7)\times 10^{-6}$
$M_\odot$\,yr$^{-1}$.

The slight asymmetric accretion column distribution between
eclipse-ingress and eclipse-egress seen by \emph{MAXI} was also found
by \emph{RXTE} and \emph{BeppoSAX} \citep{2006JApA...27..411M},
suggesting that a trailing
wake of material around the neutron star
seems to be a permanent structure in this source.

\begin{acknowledgements}
      We are grateful to the anonymous referee whose comments allowed us
      to improve this paper.
      Part of this work was supported by the Spanish Ministry of Economy
      and Competitiveness project numbers ESP2013-48637-C2-2P,
      and ESP2014-53672-C3-3-P,
      by the Vicerectorat d'Investigaci\'o,
      Desenvolupament i Innovaci\'o de la Universitat d'Alacant
      project number GRE12-35, and by the Generalitat Valenciana
      project number GV2014/088.
      This research has made use
      of MAXI data provided by RIKEN, JAXA and the MAXI team.
      JJRR acknowledges the support
      by the Matsumae International Foundation Research Fellowship
      No14G04, and also thanks the entire MAXI team for
      the collaboration and hospitality in RIKEN. The work of AGG has
      been suported by the Spanish MICINN under FPI Fellowship
      BES-2011-050874 associated to the project AYA2010-15431.
      TM acknowledges the grant by the Vicerectorat d'Investigaci\'o,
      Desenvolupament i Innovaci\'o de la Universitat d'Alacant
      under visiting researcher programme INV14-11.
\end{acknowledgements}



\end{document}